# Testing the near-infrared optical assembly of the space telescope *Euclid*

Christof Bodendorf [*a], Norbert Geis[a], Frank Grupp[b,a], Jennifer Kaminski[a], Reinhard Katterloher[a], Ralf Bender[a,b] on behalf of the Euclid Consortium
[a] Max Planck Institute for Extraterrestrial Physics (Germany).
[b] Univ.-Sternwarte München (Germany).

## ABSTRACT

*Euclid* is a space telescope currently developed in the framework of the ESA Cosmic Vision 2015-2025 Program. It addresses fundamental cosmological questions related to dark matter and dark energy. The lens system of one of the two scientific key instruments [a combined near-infrared spectrometer and photometer (NISP)] was designed, built-up and tested at the Max Planck Institute for Extraterrestrial Physics (MPE). We present the final imaging quality of this diffraction-limited optical assembly with two complementary approaches, namely a point-spread function- and a Shack-Hartmann sensor-based wavefront measurement. The tests are performed under space operating conditions within a cryostat. The large field of view of *Euclid*'s wide-angle objective is sampled with a pivot arm, carrying a measurement telescope and the sensors. A sequence of highly accurate movements to several field positions is carried out by a large computer controlled hexapod. Both measurement approaches are compared among one another and with the corresponding simulations. They demonstrate in good agreement a solely diffraction limited optical performance over the entire field of view.

## 1. INTRODUCTION

The wide-angle space telescope *Euclid* is an ESA mission, dedicated to the exploration of the nature of dark matter and dark energy [1]-[3]. *Euclid* is scheduled for launch and subsequent transfer to the second Sun-Earth Lagrangian point (L2) in 2022. From there, it will carry out a wide survey, covering 15 000 deg$^2$ (36 % of the whole sphere) of the extra-galactic sky throughout the past 10 billion years, or equivalently up to a redshift of ~ 2. During a mission life time of ~6 years, *Euclid* will map roughly 1 billion galaxies.

For this purpose, *Euclid* is equipped with a Korsch-telescope with a 1.2 m diameter primary mirror. The output of this telescope is divided by means of a dichroic beam splitter, feeding two instruments, a visible imager (VIS) and a combined near infrared slitless spectrometer and photometer (NISP) [4],[5]. Both instruments have a common field of view (FoV) of 0.53 deg$^2$. The NISP optical assembly (NI-OA) was designed and built-up at the MPE. It consists of 4 spherical-aspherical meniscus lenses with diameters of up to 170 mm (which are the largest lenses in civil space missions ever to this day) and of a slightly powered filter. The requirement of an almost diffraction-limited imaging quality over the complete wide-angle FoV in combination with the aspherical lens surfaces leads to particularly critical centring tolerances of each single lens well below 10 µm. This challenge was solved with an interferometric adjustment procedure based on multi-zonal computer-generated holograms [6]-[8]. The optical performance of the thus assembled NI-OA is tested under space-operating conditions. We show the results in terms of wavefront errors (WFE), point spread functions (PSF) and encircled energies (EE).

## 2. EXPERIMENTAL SETUP

The overall test of the optical performance of *Euclid*'s 'near infrared optical assembly' (NI-OA) under operating conditions requires an ambient temperature of 133 K (–140° C) in a vacuum chamber. The test includes independent PSF and WFE measurements at several FoV positions and thus allows the determination of further quantities like the encircled energy.

---

[*] Christof.Bodendorf@mpe.mpg.de

## 2.1 Overview

A schematic overview of the experimental setup is given in Figure 1. A photo of the actual experiment is shown in Figure 2. The entire beam path in this test-setup is *reverse* to the later *Euclid* operation, which means that the image plane of NISP is the object plane here (referred to as 'Source Plate' in Figure 1). This choice brings enormous technical advantages, as all moving parts for controlling the field positions and the detectors are mounted outside of the cryostat on a hexapod in a warm environment. The fixed source plate can thus be completely inside of the cryostat and only a single cryostat window is required for the passage of the less critical f/20 beam (see below).

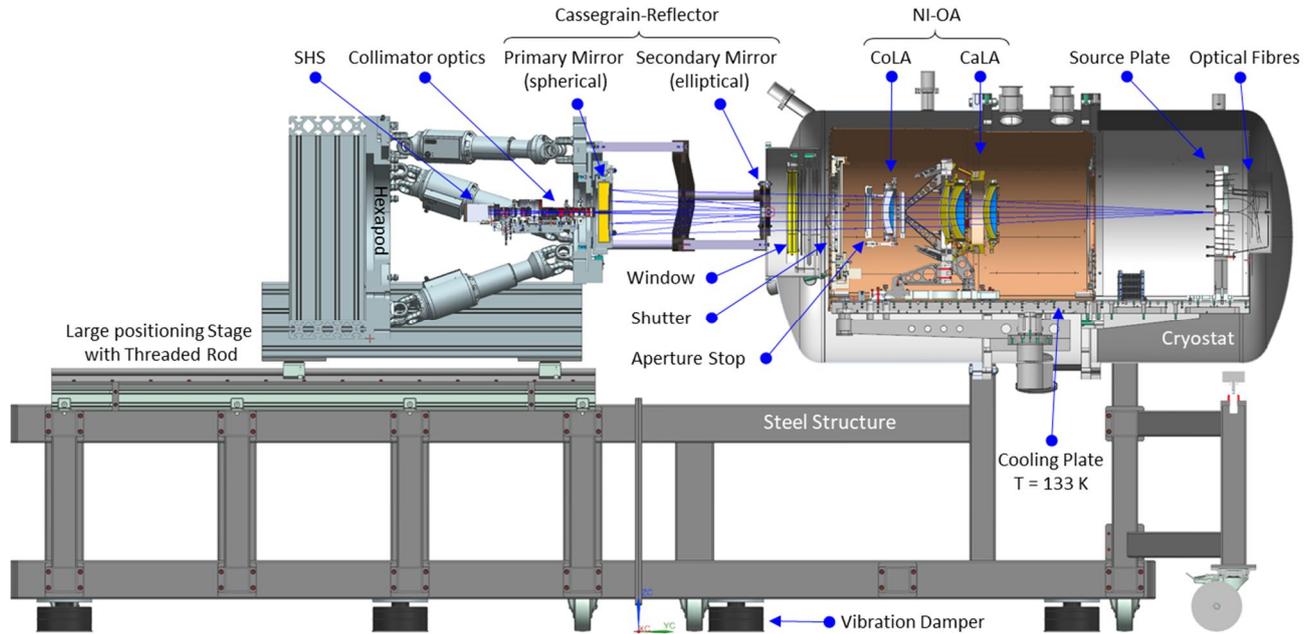

Figure 1. Setup for the optical performance test of NI-OA under operating conditions with a Shack-Hartmann sensor (technical drawing).

The object plane ('Source Plate' in Figure 1) consists of a flat fused silica plate with inserted single-mode optical fibres at 22 positions along the FoV. The fibres are guided outwards with a vacuum feed-through and connected to a light source in the warm environment.

NI-OA is the 'system under test'. It consists of a 'camera lens assembly' (CaLA) and a 'corrector lens assembly' (CoLA). For the test, they are interconnected by a monolithic invar structure dummy, which replaces the NISP structure into which CaLA and CoLA are finally integrated.

NI-OA transfers an f/10 beam from the source plate to an f/20 beam. The subsequent aperture stop, located at the position of the exit pupil of *Euclid*'s *Korsch* telescope, determines the required numerical aperture (NA) of 0.0501.

The diverging f/20 light cone leaves the cryostat through a flat fused silica window with a thickness of 30 mm and high optical surface quality.

Since the NI-OA alone is not an imaging optics, a Cassegrain like two mirror telescope (referred to as 'measurement telescope' in contrast to *Euclid*'s Korsch telescope) is used to transform the diverging beam into a converging one. The telescope has a magnification of −1, which means that the system consisting of NI-OA *and* telescope has a magnification of −2. As the telescope mirrors are installed in an invar structure, the magnification is insensitive to temperature fluctuations. This is particularly important for determining the image field curvature. The entire telescope is mounted on a large hexapod and can be moved to the different field positions with an accuracy in the range of individual microns.

Two interchangeable detector modules are available and can be mounted to the left side of the working platform of the hexapod:

- For the WFE measurement, the beam is collimated with two plan-convex lenses ('collimator optics' in Figure 1) behind the image plane of the telescope and detected with a *Shack-Hartmann sensor* (SHS). The micro lens array

('lenslets') of this sensor is positioned in the pupil plane. The detected wavefront at this position is unique and has a sharply defined edge, namely the image of the aperture stop.

- In case of the PSF measuring setup (shown later in Figure 10), an additional plano-concave 'Barlow-lens' increases the focal length and thus the magnification by a factor of three. A subsequent Peltier-cooled low noise CCD camera is used without an objective lens. Hence, the PSF is mapped directly onto the CCD image sensor with an overall magnification of – 6.0.

The complete setup is mounted on a robust steel structure, which is mechanically isolated from the environment by vibration absorbers. The vacuum pumps (not shown) are turned off during the measurements to reduce vibrations.

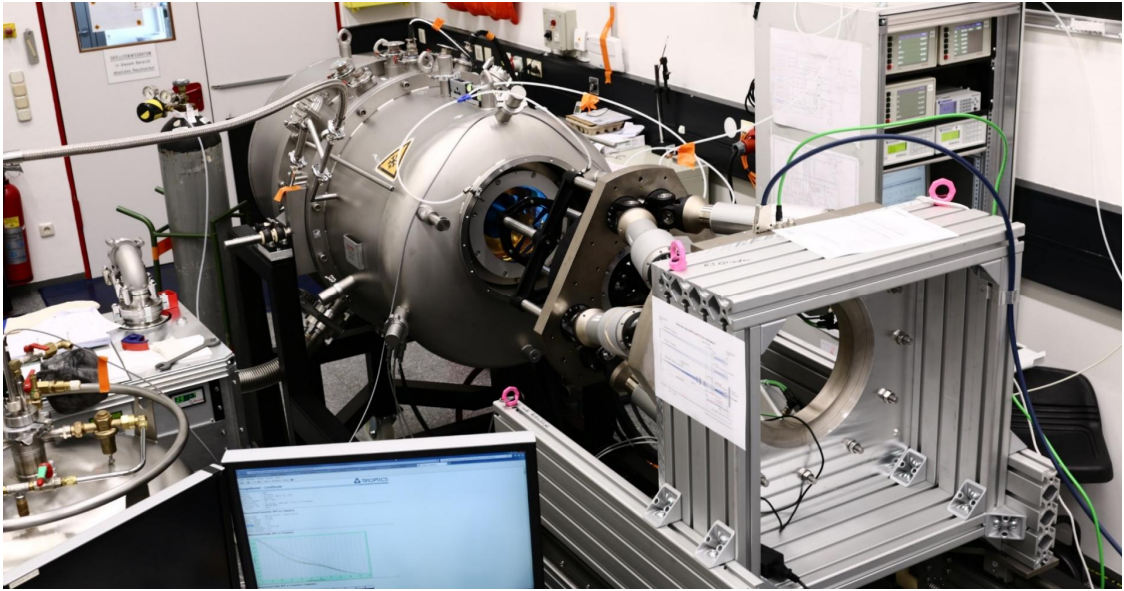

Figure 2. Laboratory setup for the optical performance test of NI-OA under operating conditions.

## 2.2 Object Plane and Illumination

### 2.2.1 Source-Plate

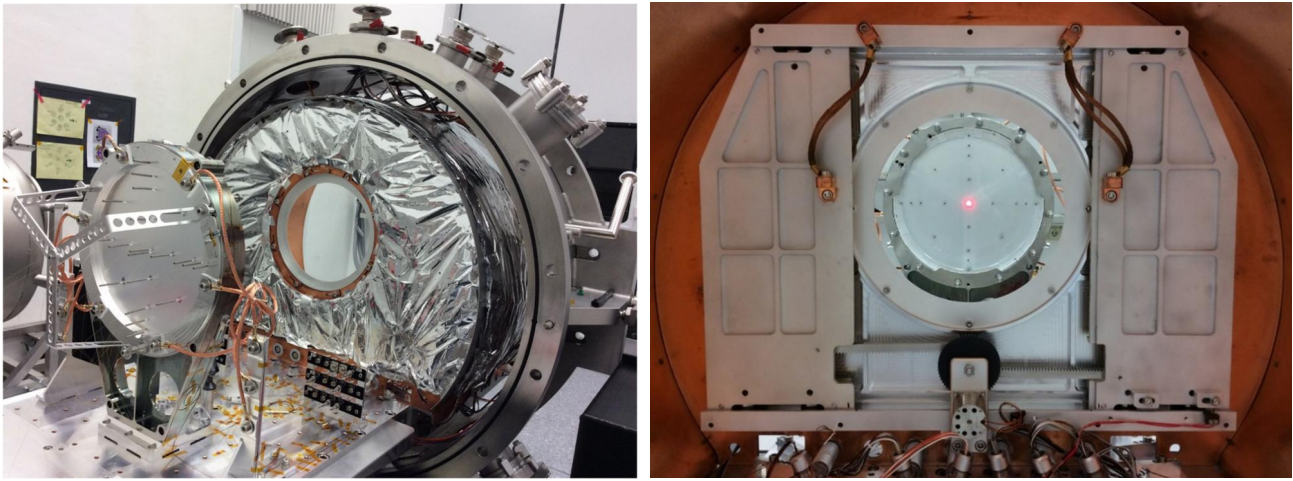

Figure 3. Source-plate with 22 single-mode optical fibres during integration into the cryostat. Left image: Rear view. The invar mount is thermally coupled to the actively cooled aluminium base-plate with several copper braids. Right image: Front view. The central fibre is turned on ($\lambda$ = 633 nm).

The position of the object plane is precisely defined by a flat fused silica plate with a diameter of 257 mm, integrated into an invar mount (Figure 3). The 'object points' are approximated by inserted and glued single-mode optical fibres at 22 positions along the FoV. The deviations from the nominal lateral positions are proven to be less than 50 µm. For each fibre, a field dependent drilling angle is applied to align the angle of radiation (which differs from the drilling angle due to refraction) with the chief ray of the field position (Figure 4). This assures a rotation symmetric pupil illumination fill, which is especially important for the PSF measurements. The inserted fibres are polished together with the substrate to achieve an accurately defined flat object plane with a peak-to-valley value of 1/3 µm as reference for the image field curvature measurements.

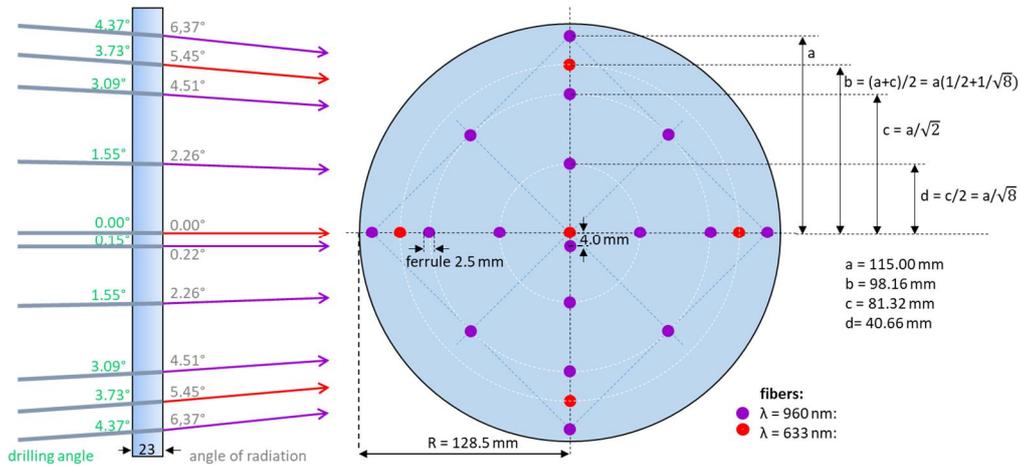

Figure 4. Positions and angles (drilling angle and angle of radiation due to refraction) of the fibres inserted in the source plate. Two different fibres for two wavelengths are used.

### 2.2.2 Optical Fibres and Light Sources

We use 17 single-mode optical fibres with a core diameter of 5.1 µm and NA 0.12 (values according to specification sheet), distributed over the FoV as shown in Figure 4. The fibres are fed with a superluminescent diode (SLED) at $\lambda = 960$ nm with a bandwidth (FWHM) of about 30 nm. This wavelength is close to the lower end of the NISP operating range (900 - 2000 nm for the photometric channel) but still observable with standard CCD sensors. The corresponding 'mode field diameter' (MFD) for $\lambda = 960$ nm is 6.5 µm. The fibre modes couple into free space as Gaussian beams [9]. According to the *beam parameter product* (BPP) $\theta \times \text{MFD} \cong \lambda/\pi$, where $\theta$ is the half convergence angle, the beam divergence due to the spatial extent of the source leads to an intensity decrease from 100% in the centre of the pupil to 55% at the edge. The ratio between MFD and the Airy diameter $D_{\text{Airy}} = 1.22 \lambda/\text{NA} = 23.4$ µm can be used as measure for the deviation from a 'point source' like behaviour [9]. Here we get MFD / $D_{\text{Airy}} = 0.28$. According to [9] section 3.4 (and Figure 25 later in this paper), this value leads to slight, but already observable deviations from an ideal point source in the PSF measurement. From this point of view, an even smaller MFD as available with 'Ultra-High NA' fibres would have slight advantages. However, the practical handling of fibre connectors and vacuum feedthroughs under cryogenic conditions for these fibres with MFDs in the order of only 2 µm is not reliable.

The 5 fibres for $\lambda = 633$ nm (Figure 4) are fed with a He-Ne laser. They are mainly used for adjustment purposes. As this wavelength is outside of the NISP operating wavelength range, the quantitative results for these fibres are of less relevance.

### 2.3 Near Infrared Optical Assembly (NI-OA)

NI-OA is the 'unit under test'; see Figure 5 for the optical design. Its task is a beam transformation from f/20 to f/10 - with almost no impact to the image quality for the complete FoV. This reduces the focal length for a compact design as well as the required area of the NISP's expensive H2GR™ (Teledyne, USA) detector. Again, in our test setup with reversed beam path, it is an f/10 to f/20 transformation. The three spherical-aspherical meniscus lenses L1, L2, L3 with a diameter of 170 mm are combined to form CaLA, while the single spherical-aspherical meniscus lenses L4 constitutes CoLA. The plano-convex filter dummy with slight spherical power is identical to the NISP *Y*-band filter, but without filter coating. *Euclid*'s aimed very high imaging quality over the complete FoV requires an extraordinary high alignment accuracy of NI-OA. The corresponding interferometric procedure is described in Ref. [6] and [8].

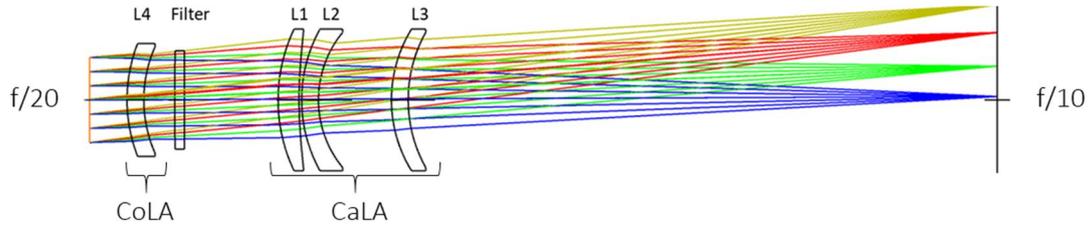

Figure 5. Optical design of NI-OA, consisting of CaLA, CoLA, and a filter. The four beams correspond to the four FoV positions of Figure 4 for λ = 960 nm in the negative y-direction.

Since our imaging quality verification test has to be performed prior to the integration of CaLA and CoLA into the NISP instrument, an invar 'dummy structure' was developed to connect CaLA and CoLA as shown in Figure 6. The alignment of NI-OA with regard to the source plate is described in [8]. For further information about NI-OA, see Ref. [4], [5], [10].

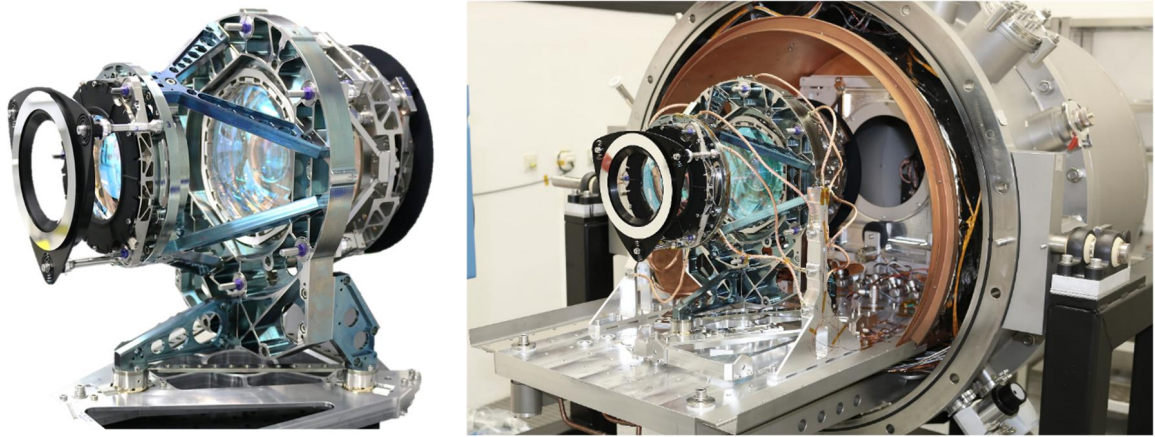

Figure 6. NI-OA, consisting of CaLA, CoLA, and a filter, mounted on an invar dummy structure. The aperture stop (mounted on the left end) defines the NA in the test-setup and is not a part of NI-OA. Left: NI-OA alone. Right: Integration into the cryostat.

## 2.4 Cryostat Window

In our test-setup, the diverging f/20 light cone leaves the cryostat through a flat fused silica window ($n_{glass}$ = 1.4509 at 960 nm) with 250 mm diameter and 30 mm thickness. This has two consequences, one in principle and one in technical terms:

- The beam displacement due to an *ideal* plane-parallel plate causes additional aberrations. Defocus is the dominating term.
- The *real* manufactured window has certain inhomogeneities and surface errors.

Concerning the first item, the defocus term has to be compensated by an increased distance to the measurement telescope of 9.33 mm (theoretical value for the ideal design and λ = 960 nm). Higher Zernike terms like spherical aberration add errors to the wavefront of well below 1/100 µm RMS, even for the worst case in the field corners, and are thus negligible.

The 'real' window (second item) was examined by a 'stitched' result of several interferometric wavefront measurements in transmission with a double-path setup. The optically used range for all FoV positions is merely the area inside of the red circle in Figure 7. For this area, the WFE RMS of 0.080 µm is almost completely dominated by a *defocus* Zernike term. The corresponding slight optical power is compensated by shifting the measurement telescope another 0.34 mm further away. This leads to an overall impact of the *real* window, which is compensated by a shift of 9.33 + 0.34 = 9.67 mm. This distance was also experimentally verified with a SHS based defocus measurement with and without window.

All other aberrations do not play a significant role, even at the corners of the FoV. The impact to the field curvature is also small. Deviations from the nominal image plane are below 11 µm, which is a tiny effect compared to the Rayleigh length ½ λ / $NA^2$ = 191 µm. Both results are confirmed by simulations.

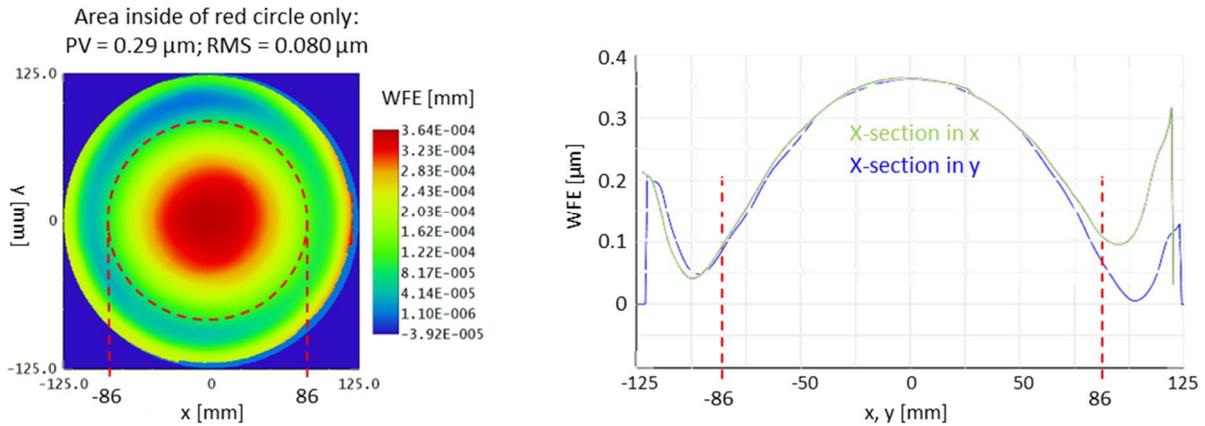

Figure 7: Interferometric wavefront measurement of the cryostat window with a double path setup in transmission. Due to the large diameter of 250 mm, a stitching technique was necessary. The optically used area in the experiment is marked by the red circle (left plot) and the dotted red lines (right plot).

## 2.5 Measurement Telescope

A two mirror Cassegrain-like reflecting telescope transforms the diverging beam into a converging one (Figure 8 and Figure 9). Its primary mirror M1 with a diameter of 190.0 mm has a spherical surface, while the secondary mirror M2 with a diameter of 32.7 mm is elliptically shaped. The mirrors are made of silver coated fused silica and assembled with an invar structure. The magnification of the telescope is –1.0.

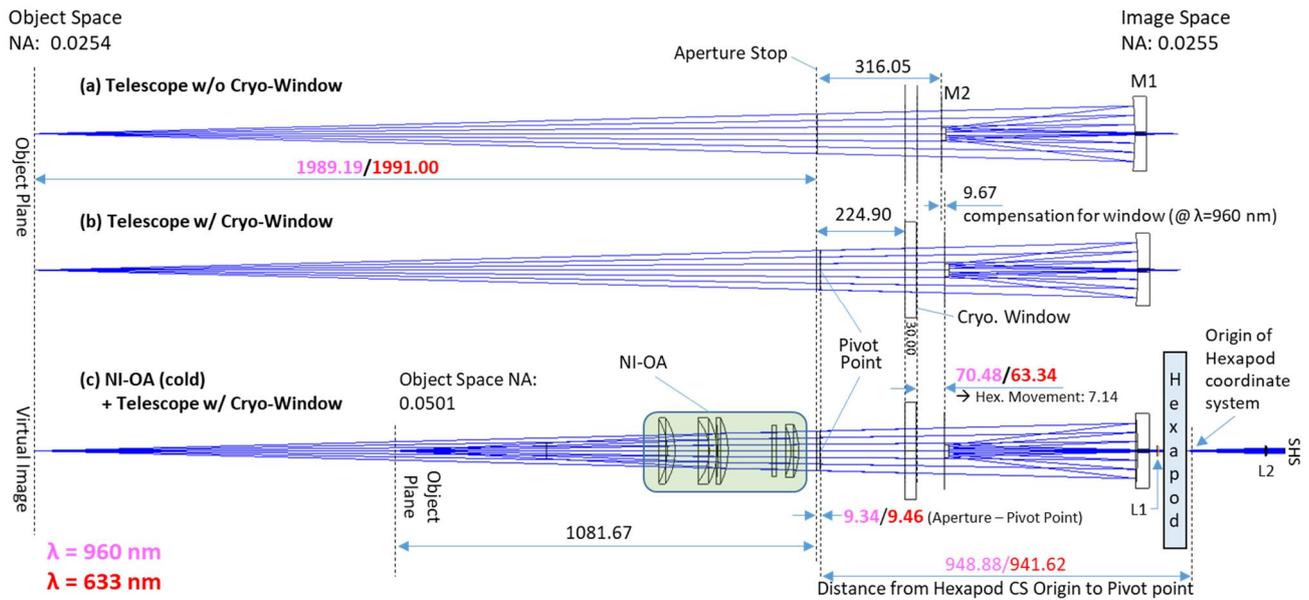

Figure 8: Optical design of the measurement telescope alone (a), with cryostat window (b) and combined with the NI-OA (c). Some distances differ for the two used wavelengths. They are printed in pink and red. All distance units are millimetres.

The key advantage of the folded telescope over a lens optics is the reduced length of the setup. This allows the telescope together with the backend optics and sensors to be mounted on a hexapod (Figure 9), to move to the different field positions. The obscuration ratio of the pupil due to M2 is 0.28, which is a bit less than the obscuration of *Euclid*'s Korsch telescope. M2 is mounted on four 'spider arms' with 1.0 mm width.

The WFE of the *nominal* telescope design is completely negligible in the field centre and $< \lambda/20$ at 960 nm for a field of 2 mm. The actually required field to map the PSF is however tiny. (The *real* telescope is investigated later in section 2.8.)

The optical design of the telescope is shown in Figure 8(a). A 'placeholder' marks the position of the cryostat window. In (b), the window is inserted and compensated by shifting the telescope to the right. The full setup with NI-OA is shown in (c). Here, NI-OA is inside of the cryostat (not shown). All three configurations share the same beam path in the telescope (apart from additional aberrations).

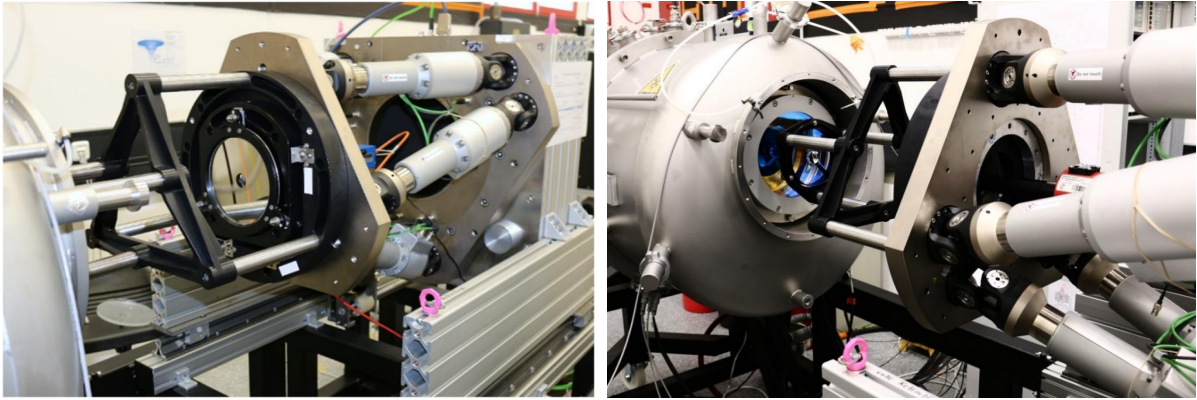

Figure 9. Measurement telescope as part of the movable 'pivot-arm' mounted on a hexapod in the lab.

### 2.6 'Backend Optics' and Sensors

The converging beam from the measurement telescope feeds two exchangeable modules: A wavefront error (WFE) measurement device and a point spread function (PSF) measurement device (Figure 10).

For the WFE measurement, we use two plano-convex spherical lenses. The first one (BL1) acts mainly as a field lens and therefore reduces the distance to the pupil plane, where the wavefront is detected. The second lens (BL2) behind the image plane collimates the beam. The combination of the lenses is also used to adjust the exit pupil diameter to 14.6 mm, which is chosen slightly smaller than the micro lens array of the SHS with a detection area of $15.1 \times 15.1$ mm. As wavefront detector, we use a 'SHSCam UHR GE' from the company *Optocraft* with a lateral resolution of $101 \times 101$ micro lenses.

For the PSF setup, a plano-concave spherical 'Barlow-lens' increases the focal length of the optical system, and therefore the magnification by a factor of 3.0. A subsequent Peltier cooled low noise CCD camera is used without objective. Hence, the PSF is directly mapped onto the CCD image sensor with an overall magnification of –6. We use the 'Bigeye G-283B' from the company *Allied Vision* with a dynamic range of 14 bit (corresponding to 16384 grey levels), resolution of 1928 × 1452 pixel and pixel size of $(4.54\,\mu m)^2$.

According to the Nyquist–Shannon sampling theorem, a *band-limited* signal with the maximum frequency $f_{max}$ has to be sampled with at least the Nyquist frequency $f_{Ny} = 2\,f_{max}$ to receive the complete information (and avoid 'aliasing' due to undersampling). The highest spatial frequency that is *transferred* from the object to the image is the cut-off frequency of the optical transfer function $f_{Cut} = 2\,NA\,\lambda^{-1}$. Assuming $f_{max} = f_{Cut}$ leads to a required sampling of at least $f_{Ny} = 2\,f_{Cut} = 4\,NA\,\lambda^{-1}$. The sampling period of the camera is 4.54 µm (equal to the pixel size) which corresponds to a spatial camera sampling rate of $f_{Camera} = (4.54\,\mu m)^{-1}$. This yields an *oversampling* of $f_{Camera}/f_{Ny} = \lambda\,(4 \times 4.54\,\mu m\,NA_B)^{-1} = 6.3$. Here, $NA_B$ is the NA after the Barlow lens (Figure 10), given by $NA_B = 0.0501/6$ with the magnification of factor 6 from the object to the image on the camera sensor and with $\lambda = 960$ nm.

The oversampling of factor 6.3 due to the optical magnification with the Barlow lens might seem superfluous, as in principle the full information is already there with Nyquist sampling. It has however several advantages: The sampled PSF is already well spatially resolved and an interpolation with a reconstruction filter is not required. (We resolve the Airy diameter $D_{Airy} = 1.22\,\lambda\,NA^{-1}$ with 31 pixels, while pure Nyquist sampling would result in only 4.9 samples) Variations in the pixel-to-pixel sensitivity are average out e.g. in the encircled energy and a flat-field correction is usually not required. The increased number of pixels per camera frame (factor 40) improves the 'effective dynamic range' of the camera and reduces the measurement noise.

And finally, a real PSF doesn't have to be band-limited at all, as higher frequencies than the optical cut-off frequency can appear due to aberrations (without reference to the object). This is an additional argument for 'oversampling' with regard to the optical cut-off frequency.

On the other hand, every additional lens introduces wavefront errors. The simple spherical lens surfaces add by design a tiny spherical error, which is however completely negligible. The lens surfaces are tested interferometrically and of very high quality. (We chose the best of a whole bunch.) However, the dispersion of these lenses leads to the necessity to compensate the position of the telescope for the two different wavelengths by 7.14 mm, see Figure 8.

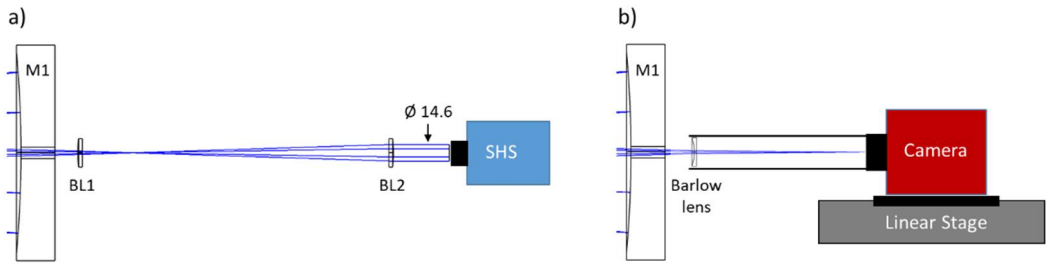

Figure 10. 'Backend' of the measurement optics. Two exchangeable modules are available: The WF-measurement module (a) with SHS and the PSF measurement module (b) with camera on a linear stage.

## 2.7 Mechanical Motion Sequence

The telescope, the 'backend optics' and the sensors constitute the movable 'pivot-arm', mounted on the hexapod (Figure 1, Figure 2, Figure 9). The optical design with 4 field positions along one diagonal of the FoV is shown in Figure 11. The mechanical motion sequence consists of a *rotation* and a subsequent *translation*. The centre of the rotation (pivot point) is the intersection point of the chief rays from the different fields, *as seen from behind the fixed cryostat window*. The intersection point of the chief rays is (by definition) the middle of the aperture stop. However, *from behind the window*, these rays appear to intersect at the middle of the virtual image of the aperture stop. Therefore, the system has *two* pivot points: The rays on the left side of the cryostat window are rotated around the middle of the aperture stop, while the rays on the right side of the window are rotated around the middle of the virtual image of the aperture stop as seen through the window. This is the result of the beam displacement by a plane-parallel plate. (The rotation is followed by a shift, see below.)

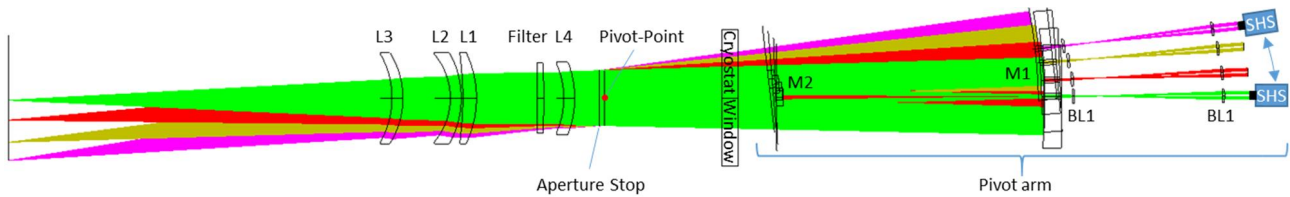

Figure 11. Optical design of the WFE measurement setup with 4 field positions. The pivot arm must be rotated about the pivot point, which is in the middle of the virtual image of the aperture stop as seen through the cryostat window.

The distance *s* from the aperture stop to this virtual pupil can be calculated with basic geometric considerations as shown in Figure 12. The angle between chief ray and optical axis in the aperture stop is denoted as α:

$$s = d \left[ 1 - \frac{\tan\left(\arcsin\left(n_{air}/n_{glass} \sin a\right)\right)}{\tan a} \right] \qquad (1)$$

Since the image viewed through a plane-parallel plate is aberrated, *s* has a slight angle dependence, shown in the inserted plot of Figure 12 for the two available wavelengths. So strictly speaking there is not one exact pivot point. This is however a negligibly small effect. We choose $s = 9.34$ mm for $\lambda = 960$ nm and $s = 9.46$ mm for $\lambda = 633$ nm. These are averages over the swing angles that yield the best simultaneous results for all available fields according to the simulation.

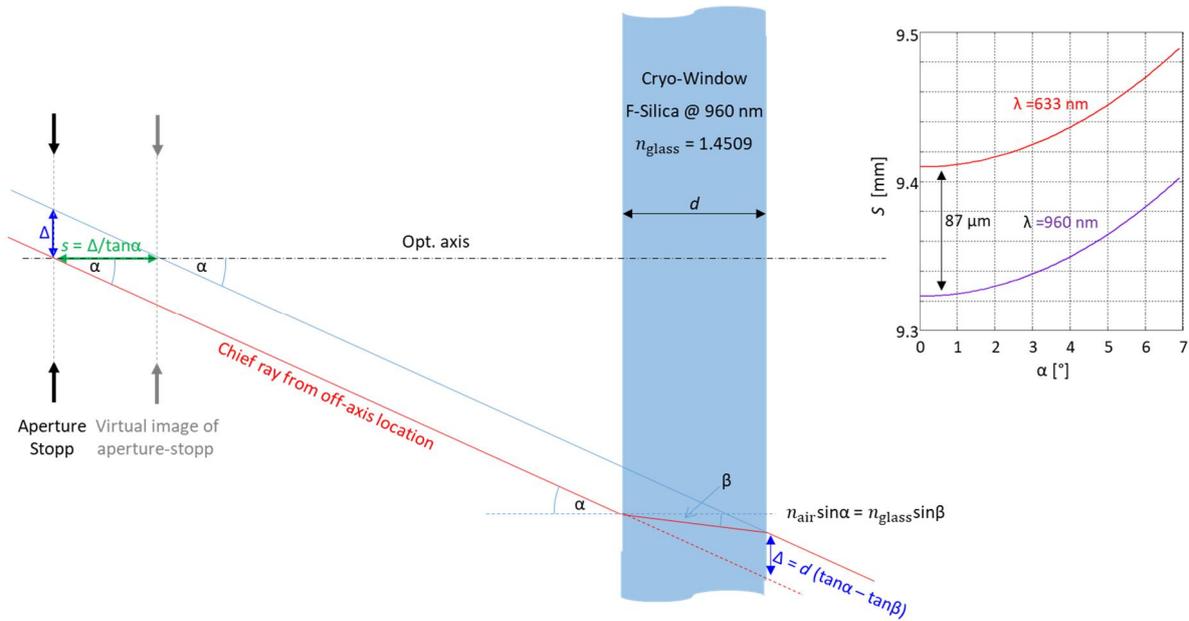

Figure 12. Calculation of the pivot point location according to Eq. (1). (Sketch not to scale.)

The *rotation angles* of the pivot arm are equal to the angles of the chief rays from the different fields in the aperture stop. This way the chief rays coincide always with the optical axis of the pivot arm. Subsequently, the entire pivot arm performs a translational motion along its axis towards the window to the nominally best focus position, which is defined by the object plane of the flat source plate and the nominal NI-OA. The angles and shifts for the four field positions at $\lambda = 960$ nm along a diagonal of the field are numerically and graphically put together in Figure 13. The entire motion sequence for all fields is software controlled and executed by the hexapod with high precision.

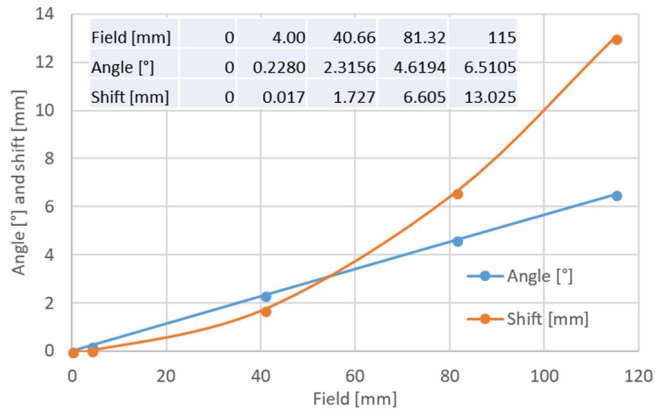

Figure 13. Nominal angles and shifts of the pivot arm for the four field positions with $\lambda = 960$ nm along a diagonal of the field.

This arrangement ensures that the measuring optics on the pivot arm is only utilised on-axis, while the FoV of the NI-OA is tested. With the tip/tilt evaluation of the SHS, the exact angles can be fine tuned with an accuracy in the order of single angular seconds. Therefore, the aberrations of the measurement optics remain unchanged for all NI-OA field locations.

Strictly speaking, this statement disregards that the cryostat window cannot be rotated together with the pivot arm and therefore adds systematic field dependent aberrations. A movable window would technically be extremely costly, as the complete atmospheric pressure strains the window. Fortunately, the additional WFE RMS contribution from the centre to the field corner due to the window is $< \lambda/200$ and may be neglected, see Figure 15 later on. The substantial impact of the window can thus be summarized as an overall focus term and a shift of the pivot point according to Eq.(1).

## 2.8 Imaging Quality of the Measurement Optics

Now that we have discussed the individual components, the optical quality of the measuring optics as a whole is tested. We first investigate the SHS-based wavefront measurement and then turn to the camera-based PSF measurement.

### 2.8.1 SHS based Wavefront Measurement

The 'out of the box' accuracy of Shack-Hartmann sensors (SHS) [11], [12] is limited mainly due to fabrication tolerances of the micro-lens array. Highly accurate is however the *repeatability* of the measurements. Therefore, a *calibration step* with a reference wave-front is performed to determine the actually relevant 'zero' positions of the array of spots [9].

The measurement of a plane wave with very low aberrations is shown in Figure 14(a). Here, the error is actually not dominated by the wavefront, but rather by the deviation of the spot array from an ideal equally spaced grid. It is a unique 'footprint' of our sensor, used to calibrate its intrinsic errors.

The optical performance of the pure measurement optics in the absence of NI-OA is evaluated with the setup according to Figure 8(b). The cryostat window is already included. The resulting wavefront error is shown in Figure 14(b). This measurement refers to the plane wave calibration in (a). The WFE RMS value of about $\lambda/45$ at $\lambda = 960$ nm is far below the *Maréchal criterion* that demands a WFE RMS smaller than $\lambda/14$ to be considered as *diffraction limited* [13]. This is an excellent result, as desired for a good measurement device. The main error contribution is a spherical aberration (not shown).

The evaluation in Figure 14(c) is based on the same measurement as in (b), but *without* reference to a calibration measurement. It is of particular importance because this is our *reference for all wavefront measurements of the NI-OA*. It consists of the intrinsic errors of the SHS as well as the optical errors of the measuring device. Therefore, both errors of different origin are considered and compensated. In other words: *The aberrations of NI-OA are observed as deviation from this distribution*. This procedure makes the setup sensitive even to very small aberrations.

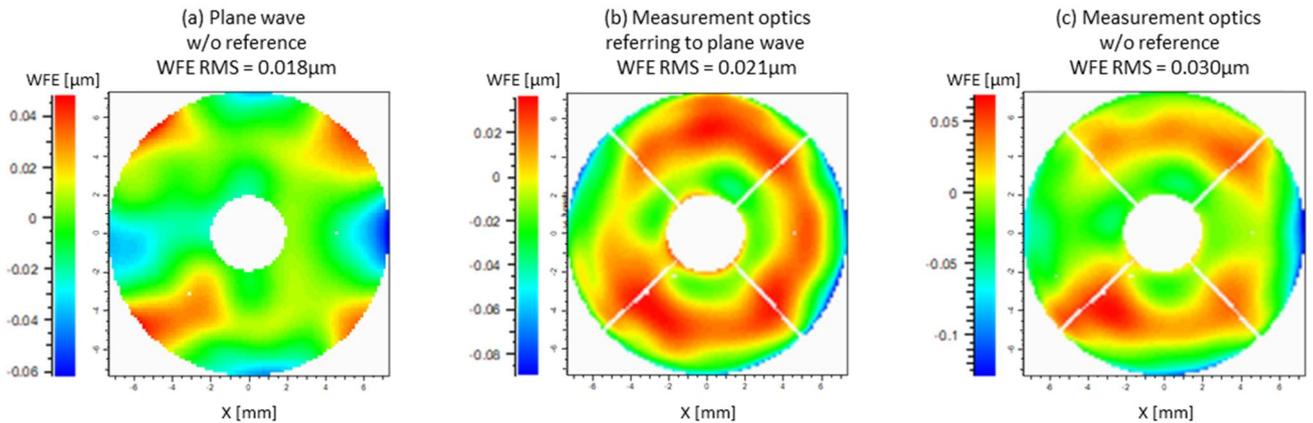

Figure 14. WFE measurements @ $\lambda = 960$ nm with a SHS. Piston, tip, tilt and defocus are removed. (a): Characterization of the SHS with a highly accurate plane wave. (b): WFE of the measurement optics, referring to the plane wave in (a). (c) Same measurement as in (b) but with the unreferenced SHS. Correspondingly, this distribution is the reference for the optical test of NI-OA.

The temporal stability of the setup was tested with continuous measurements over several days. No relevant drift of the reference wavefront in Figure 14(c) was observed. However, the distance from the source to the sensor of more than 4 meters in air increases the measurement noise due to air turbulence. This noise is averaged out over 32 single frames and reduces the standard deviation of the WFE RMS to 0.002 μm.

The aberrations - again with the setup according to Figure 8(b) - along a diagonal of the FoV are shown in Figure 15. The field dependence due to the oblique beam path through the fixed window is hardly observable. Here each dot corresponds to an average over only 8 single wavefront measurements and the impact of air turbulences to the noise becomes more significant.

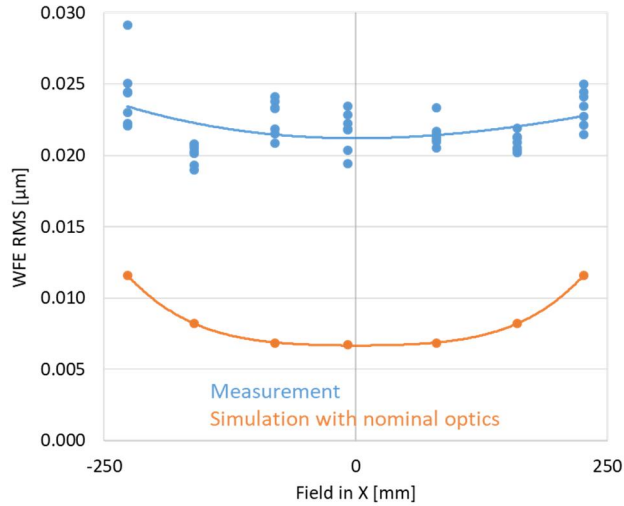

Figure 15. WFE RMS measurement along a diagonal of the FoV @ λ=960 nm. The oblique beam path through the cryostat window has only marginal impact to the optical performance.

### 2.8.2 Camera-based PSF measurement

The PSF measurement requires an accurate determination of the best focus position. For this purpose, we scan the camera (with Barlow-lens) on a motorized linear stage along the optical axis (Figure 10) and determine the maximum of the modulation transfer function (MTF), which we define as the best focus position.

At best focus, the PSF is determined as the difference between a *dark frame* with the source turned off and a '*source frame*', both with identical exposure time [9]. Within the framework of this paper, we *always* rescale the spatial expansion of the observed PSF *as it would appear on the NISP sensor* with NA = 0.0501. This means a reduction by a factor of 6.0 (see section 2.6).

Figure 16 shows a comparison between the simulated PSF with the nominal optics (a), a PSF derived from the SHS measurement (b), and a PSF from the camera measurement (c). Plot (b) is derived from the WFE in Figure 14(b). We use a logarithmic scale over four orders of magnitude. This is a reasonable adaption to the dynamic range of the camera with $2^{14} \approx 10^{4.21}$ grayscale steps. The depicted spatial range is 250 µm.

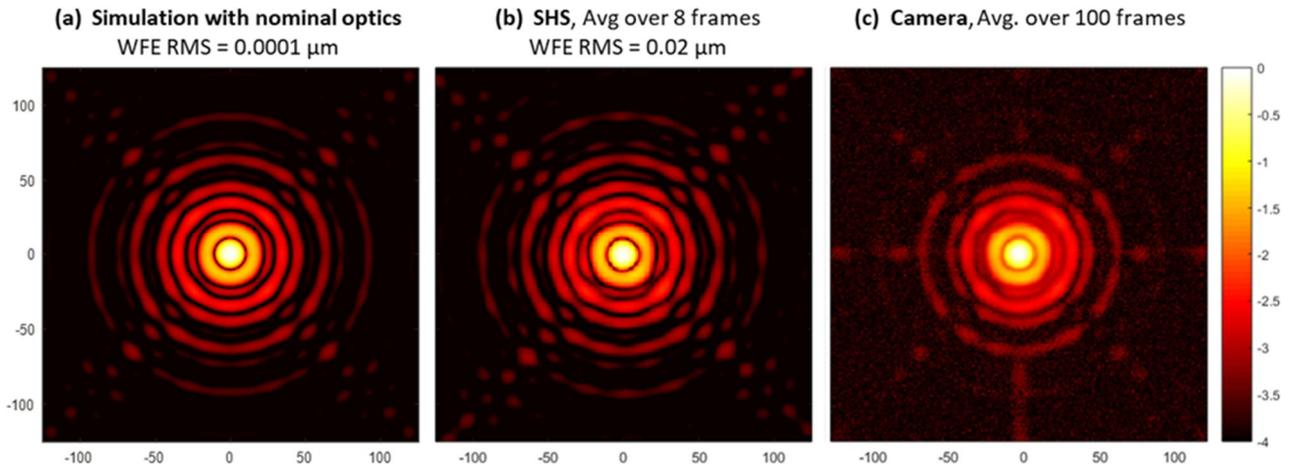

Figure 16. PSF the of the measurement optics @ λ=960 nm. Log. scale. (a): Simulation of the nominal optics. (b): Derived from the wavefront measured by the SHS. (c): Camera measurement with subtracted dark frame.

The structure of the PSF is more complex than the typical Airy function by a circular aperture. The inner obscuration of the pupil due to the telescope mirror M2 (Figure 8) leads to irregular intensities and spacings of the diffraction rings. The four 'spider arms' of M2 break the rotation symmetry and lead to a diffraction phenomenon of diagonal light spots. The slight additional horizontal and vertical light tracks that only appear in the camera measurement probably arise due to interference at the sensor pixels that act as optical grating in combination with back reflections at the camera window. The vertical line at the bottom is moreover intensified due to readout artefacts from the bright centre of the PSF.

**2.9 Adjustment of Pivot Arm in front of Cryostat**

The complete setup according to Figure 1 can be considered as consisting of two parts: The first part is the fixed group inside the cryostat, which is optically adjusted with highly accurate CGH-assisted interferometric methods [6], [8]. The second part is the pivot arm, mounted on the hexapod. It is adjusted with optical and mechanical methods based on an alignment telescope and a coordinate measuring machine and yields high optical quality as demonstrated in the last section.

Both parts are precisely merged in front of the cold cryostat with the help of information gained from the SHS: The pupil is centred in the sensor plane and tip, tilt and defocus are simultaneously adjusted to zero. This adjustment is done by moving the hexapod. Subsequently, the coordinate system of the hexapod is set to 'zero' and the required pivot point is set in this system. The setup is ready for measurements as soon as the nominal operating temperature of 133 K is stable.

## 3. IMAGING QUALITY OF THE NEAR-INFRARED OPTICAL ASSEMBLY

Now, we present the measurements with the complete setup as discussed in the previous chapter at the NISP operating temperature of 133 K in vacuum.

**3.1 Wavefront Measurements**

**3.1.1 Raw image and WF reconstruction with the SHS**

Figure 17 shows a raw image of the SHS at three zoom levels, taken at the FoV centre of the NI-OA. The spots of the lenslet array in the pupil are well resolved. The logarithmic intensity scale shows clearly the central peak and the first diffraction ring. The centroids of the spots are the basis for the reconstructed wavefront in Figure 18(a). All wavefronts in this chapter are calibrated with the reference measurement from Figure 14(c). Therefore, the aberrations of the measurement optics are compensated and the quantitative results refer solely to NI-OA. This is a necessary prerequisite for measuring such small wavefront errors with a SHS.

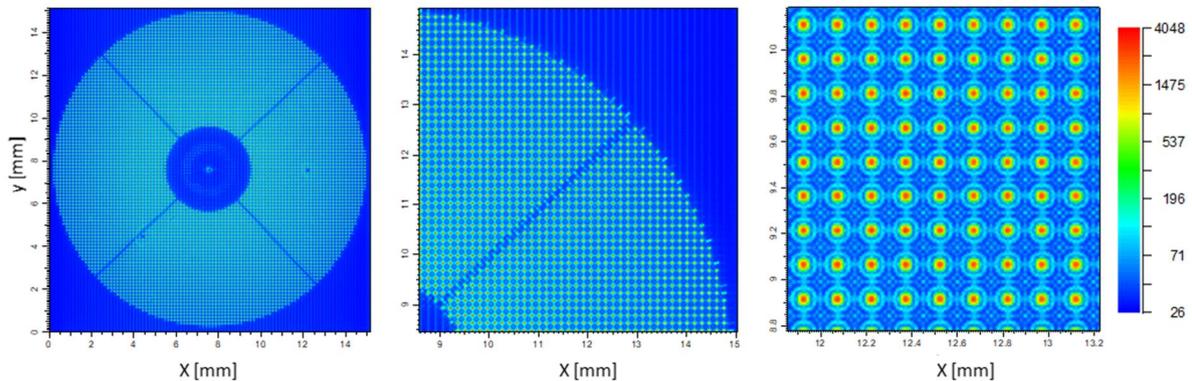

Figure 17. Raw image recorded with the SHS in the pupil plane at $\lambda = 960$ nm. The spot array is the basis for the reconstruction of the wavefront. Two spots are missing because the corresponding micro-lenses are intentionally blackened for referencing purposes. Left: Complete sensor area. Middle and right: Detail enlargements.

The reconstructed wavefront in Figure 18(a) is based on a polynomial fit (*linear least squares method*, LSQ, [14]) of 16[th] order. The 'residual' in Figure 18(b) shows the deviation between directly measured spot positions and the corresponding predictions from the reconstruction. It is a consistency check that e.g. indicates whether the polynomial form is adequate to describe the wavefront. If this is the case - as here - then it basically gives an impression of the random measurement noise, which is not included in the fit and thus filtered out in the WF.

Figure 18(c) shows the intensity distribution in the pupil. These measurements are not accurately calibrated and give rather qualitative than quantitative results. Nevertheless, the Gaussian intensity decline with radius (section 2.2.2) is well observed. Slow variations in the scale of the single micro lenses (diameter 150 µm), have no observable impact to the determined WFE. The distribution is well centred, indicating that the angle of radiation from the fibre is correct (section 2.2.1). This is true for all fields. It is actually more important for the camera PSF measurement than for the SHS measurement, which also tolerates a certain degree of asymmetric pupil fill without significant impact to the WFE result.

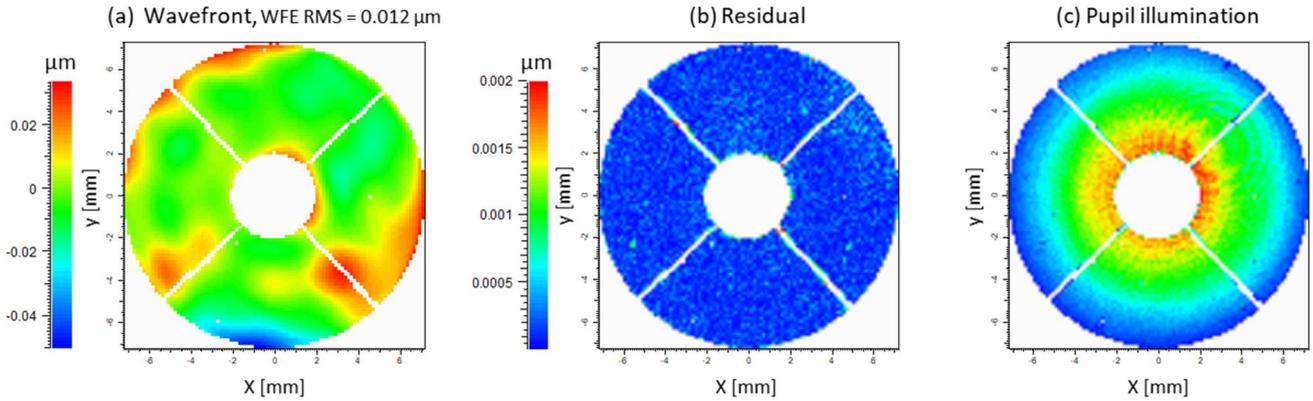

Figure 18. (a): WFE of the NI-OA measured with the SHS at λ = 960 nm in the centre of the FoV. (b): Residual between spot locations as predicted with the reconstructed WF and the measured values. (c) Gaussian intensity distribution in the pupil (qualitative measurement).

### 3.1.2 Image Quality in terms of Wavefront Error

We performed wavefront measurements at the FoV positions indicated in Figure 4. The RMS values of the wavefront errors for λ = 960 nm after subtraction of piston, tip tilt and defocus are plotted in Figure 19 (all three diagrams have the same *y*-scaling). In (a) the field positions are located along the second white circle drawn in the source plate in Figure 4 with radius 'c'. (One vacuum feedthrough at 315° was damaged and could not be connected.) In the last position of this plot, we display additional measurements in the field centre, which were taken several times in between. In (b) and (c) the positions are chosen along the diagonals in *x* and *y*. For each position, five WFEs are determined, whereby each WFE is averaged over 64 frames of the SHS. This statistic is necessary to reduce the disturbance induced by air turbulences, especially due to the convection airflow at the (colder) cryostat window.

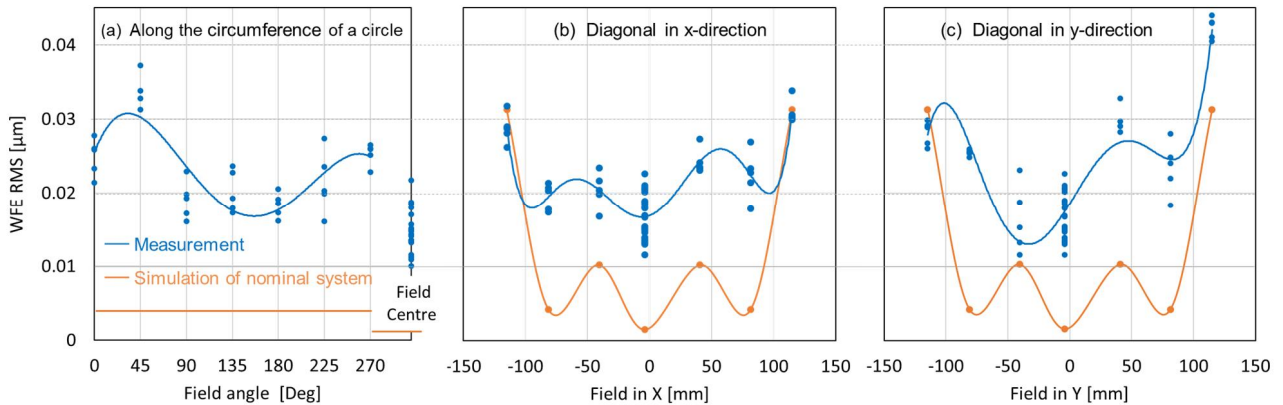

Figure 19. WFE RMS at 16 locations in the FoV of the NI-OA. All measurements are well below the diffraction limit. The lines are just polynomials to 'guide' the eyes.

The typical WFE RMS values are in a range between λ/60 and λ/30, which is better than the so-called diffraction limit of λ/14 = 0.069 µm [13] by a factor of 2 to 4. The worst case at one single corner in the order of λ/22 is still far below the diffraction limit. This applies at λ = 960 nm close to the lower end of the NISP wavelength range. The corresponding values for the upper end at λ = 2000 nm are correspondingly even better. The main contribution to the WFE RMS already comes from the low Zernike terms astigmatism, coma, trefoil, and spherical aberration (not shown).

### 3.1.3 Field of View Curvature

In the previous SHS evaluations, the defocus Zernike term was removed (together with piston, tip and tilt). Now, we evaluate the spatial position of the best focus along the chief ray, again at several locations across the FoV, to determine the *field curvature*. For this purpose, the defocus is the essential observable. The defocus Zernike coefficient is (for a sufficiently small range) a linear function of the position along the chief ray with a slope of 0.093 µm/mm. This value was determined by measuring 'focus ramps' and confirmed with simulations.

However, an additional effect must be taken into account: Just before the measurements, a shutter (radiation shield) behind the cryostat window opens up. From that moment on, the thermal radiation from outside warms the optics as well as the lens barrels. This causes a time-dependent focus drift that must be corrected. Other Zernike terms are significantly less affected.

This leads us to the following measurement sequence: The hexapod reaches successively the nominal field positions. This happens with a positioning accuracy in the range of individual micrometres. The wavefront at each position is measured several times. After a sequence of three field locations, the hexapod jumps back to the centre of the field and repeats this measurement. We thus obtain five intermediate measurements at the field centre for different times during the measuring program. The corresponding Zernike defocus terms are plotted over time in Figure 20(a). We assume a linear drift, given by the fit, which we use to correct the field measurements at the corresponding times. The thus determined and corrected defocus values over the FoV are converted to the spatial deviations from the nominal best focus positions. As these values are determined on the f/20 side, they are divided by the longitudinal magnification factor of 4 to the f/10 side of the sensor plane of the NISP. The result is presented in Figure 20(b) and (c) for the two diagonals of the FoV.

The deviations from the ideal position are in the order of 50 µm, while the spread of the measurements is almost in the same order of magnitude. This value is small compared to the Rayleigh length $½ \lambda / NA^2 = 191$ µm and therefore uncritical. A small remaining overall *tilt* [red lines in Figure 20(b) and (c)] may originate in a tilt of the source plate away from the normal to the optical axis. This can happen during the cooling process. A large centring error of the pivot point would have the same effect, but can be excluded to be the primary cause.

The slightly increasing deviation from the centre to the corners can be assumed to be a real property of the NI-OA, although it is not much larger than the spread of the measurements. A rather large placement error of the pivot point along the optical axis would have a very similar effect, can however be ruled out at the required magnitude.

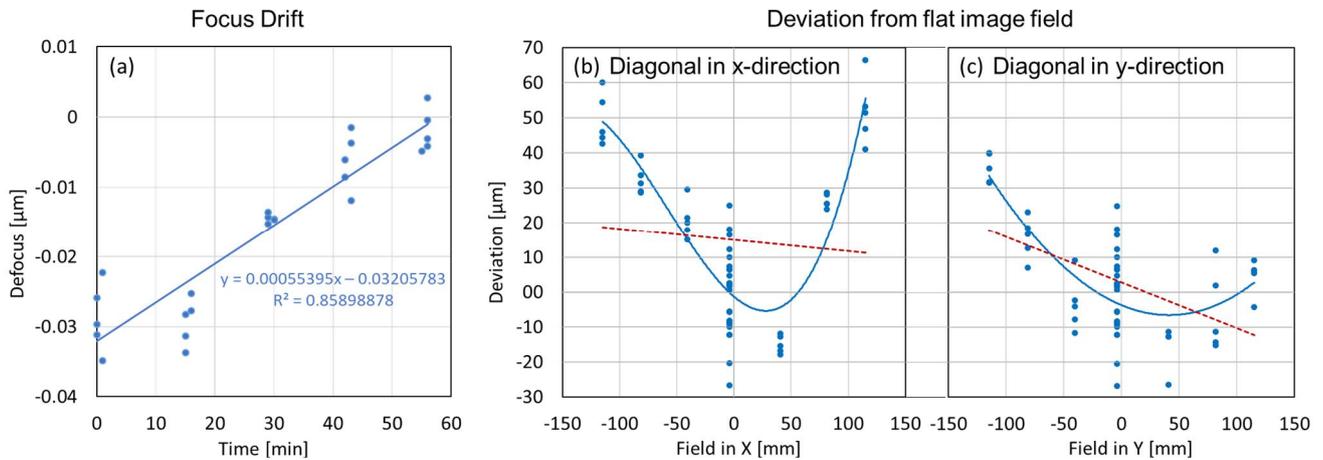

Figure 20. (a) Drift of the focus Zernike term as function of the measurement time. This drift is compensated for the calculation of the image field curvature. (b) and (c) Deviation from a flat image field on the NISP sensor plane as function of the field along the two diagonals.

### 3.2 Camera-based PSF Measurements and comparison to Wavefronts

The PSF is determined as described in section 2.8.2. The results for $\lambda = 960$ nm are shown in the logarithmically scaled plots of Figure 21. A simulation with the nominal design is inserted on the bottom left (highlighted in green). A sketch of the source plate with the corresponding field positions is on the bottom right.

The PSFs at the four corners of the FoV show an additional diffraction pattern. The reason for this disturbance is a *vignetting* due to a *rectangular* stray light baffle, adapted to the rectangular FoV and integrated in NI-OA. Unfortunately, the FoV for the tests had to be defined differently, namely rotated by 45° to lie inside of the accessible range of motion of the hexapod. Since the baffle could not be rotated, vignetting was unavoidable. See also Figure 23(d) further below. The normal operation of NISP is of course not affected. Apart from this, the shape of the PSF is extremely uniform throughout the field - as to be expected from the wavefront measurements in the previous section.

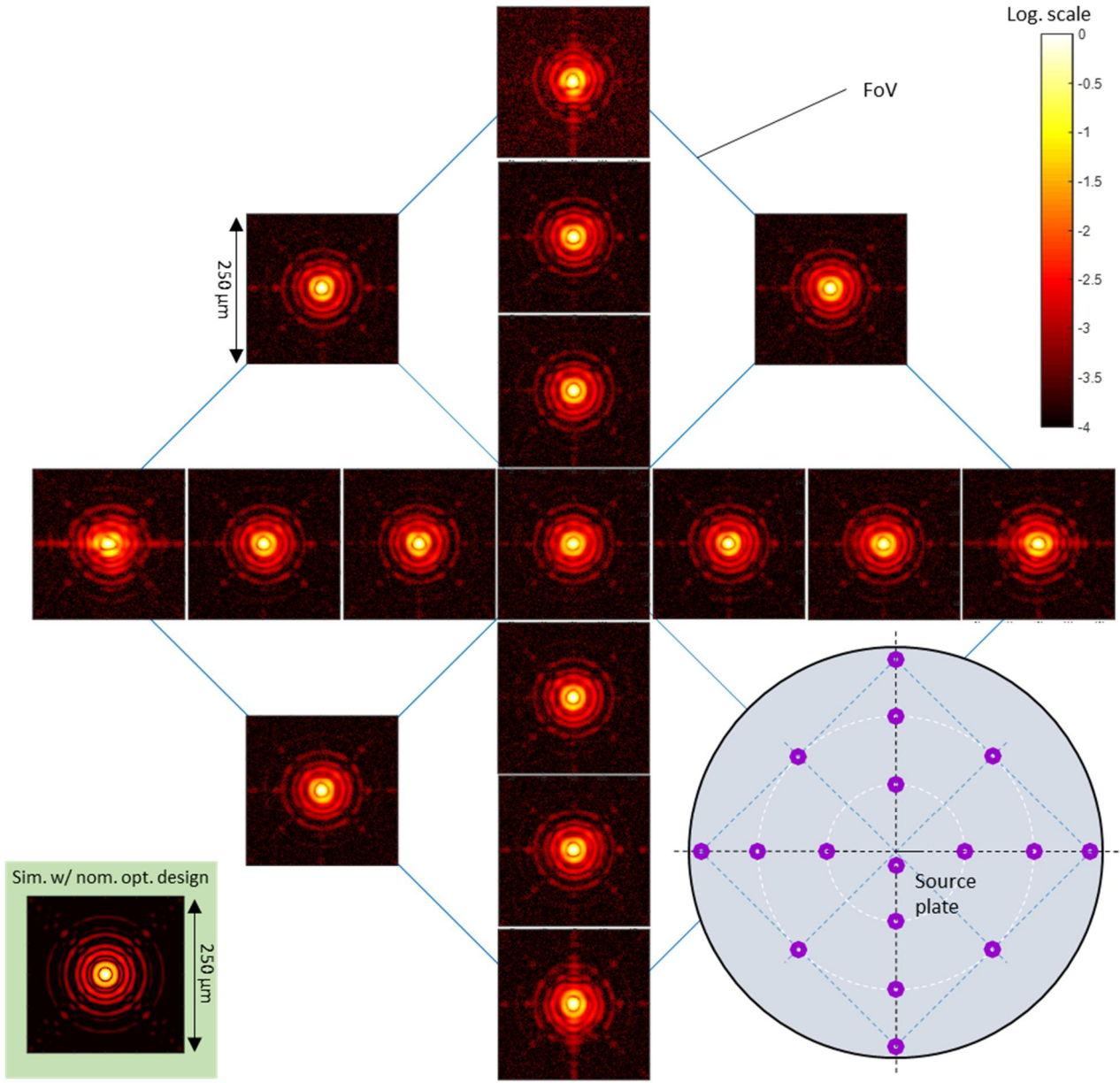

Figure 21. PSF of NI-OA at 16 image field positions; logarithmic scale over 4 orders of magnitude. The size of the images is 250 µm. A sketch of the source plate with the FoV locations is inserted on the bottom right. A simulation with the nominal design is shown on the bottom left (green highlighted).

### 3.2.1 Comparison of the PSF without and with NI-OA

In contrast to the SHS, which is calibrated to a reference wavefront, a mathematically exact distinction between the aberrations of the measurement optics and the 'unit under test' is not possible with the PSF approach. Nonetheless, the spots without and with NI-OA according to the setups in Figure 8(b) and (c) can be compared directly, as shown in Figure 22(a) and (b). [See section 2.8.2 for the spatial scaling.] From the SHS results, we know that the WFE RMS of the measurement setup alone and of the NI-OA alone are both in the order of 0.025 µm. The entire setup has therefore an RMS in the order of $\sqrt{2}$ 0.025 µm = 0.035 µm. So we are dealing with clearly diffraction limited PSFs. The difference between Figure 22(a) with RMS about 0.025 µm and Figure 22(b) with RMS about 0.035 µm is accordingly almost not recognisable in this representation. In other words: The NI-OA works so well, that we do not see any significant impact to the PSF. The slight horizontal and vertical light tracks are non-existent in the ideal PSF. As already discussed in section 2.8.2, they appear independent of NI-OA in our setup. Figure 22(c) is just a different visualization of the PSF in (b). The linear plot gives a good impression of the small relative intensities of the diffraction rings in comparison to the central peak.

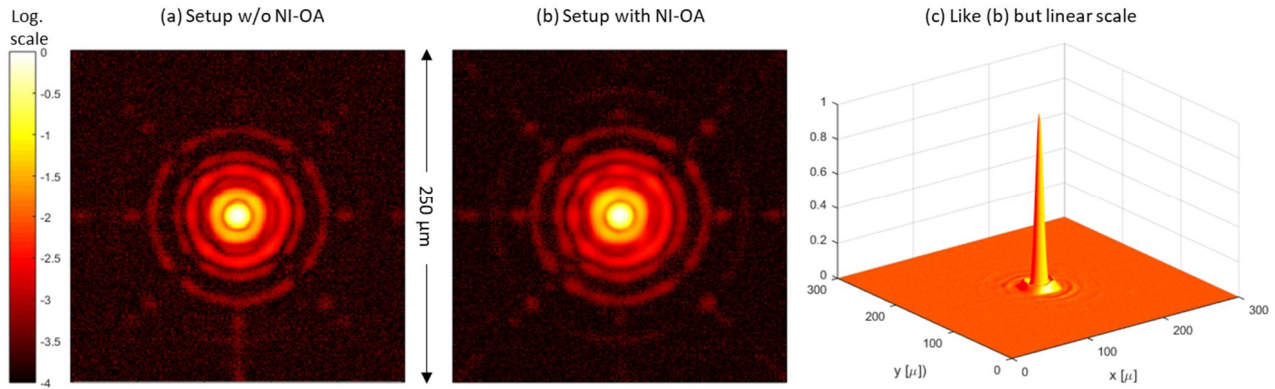

Figure 22. PSF in the field centre for $\lambda = 960$ nm. (a) and (b): Logarithmic intensity scale over 4 orders of magnitude. The difference between the pure measurement setup without NI-OA in (a) and the complete setup with NI-OA in (b) is almost not recognisable in this representation, because both are clearly diffraction limited. The linear 3D-plot in (c) gives a good impression of the tiny relative intensities of the diffraction rings in comparison to the central peak. It is just a different visualization of the data in (b). These are highly resolved images. The NISP detector system (H2RG™ detectors) has a much larger pixel size of $18 \times 18$ µm$^2$ and would resolve plot (a) and (b) with only $14 \times 14$ pixels.

## 3.3 Camera versus SHS

A comparison between SHS and Camera measurements is possible e.g. at the level of the PSF. For this purpose, we derive the SHS-based PSF as the square of the absolute value of the 2D-Fourier transform of the pupil function, which is reconstructed from the SHS measurement. The pupil function contains phase *and* intensity information. The intensity distribution [see Figure 18(c)] is however essentially given by the Gaussian beam from the spatially coherent fibre mode and *not* a property of the system under test (section 2.2.2). For this reason, we set the intensity in the pupil to a constant value (this can be done in the evaluation software) to calculate the PSF. This corresponds to the property of a mathematical point source. It is an elegant way to get rid of the impact of the finite source extent.

However, the finite object size (mode-field-diameter and according Gaussian pupil fill) affects the camera PSF measurement. It is investigated in Figure 25 and actually almost unobservable in terms of the PSF, but has a slight influence on the EE calculation.

A comparison between the thus calculated SHS-based PSF and camera-PSF in logarithmic scale is put together in Figure 23. The top row (a) shows the simulated WF and PSF for the nominal design. Here we make a small additional distinction: The 'spider arms' of the telescope mirror have a physical width of 1.0 mm. This agrees with the actual view of the camera. The corresponding simulation is shown in the upper right plot and can be compared with the camera measurements at different fields in the column below.

However, the SHS 'sees' these 'spider arms' with the resolution of the lenslet array of 150 µm. Two successive spots in a row and in a column are shadowed and thus below the chosen threshold of the evaluation algorithm. This leads to an 'effective' spider width of $\sqrt{2} \times 150$ µm $\times 101/14.6 = 1.5$ mm. Here, $101/14.6 = 6.9$ is the pupil magnification. The correspondingly simulated PSF (top row, middle plot in Figure 23) shows slightly more intensity in the diffraction patterns in the diagonals. This leads to a very good agreement with the PSFs in the middle column beneath, which is derived from the SHS based wavefronts in the left column.

The impact of the spider to the camera PSFs is slightly smaller – as predicted by the corresponding simulation with the physical spider width of 1 mm. Another difference in the two measurement approaches is the noise level: The huge intensity range of 4 orders of magnitude leads to considerable noise in the camera measurement for the very low intensities. That is why we show averages over 40 individual camera images, which in turn results in a slight blur and loss of contrast.

The truncated wavefront due to vignetting (section 3.2) in the last row (d) of Figure 23 has a considerable impact to the PSF in terms of additional 'jets' in vertical direction (slightly rotated). This effect is evident in both measurement approaches with good consistency. It was moreover confirmed by simulations (not shown).

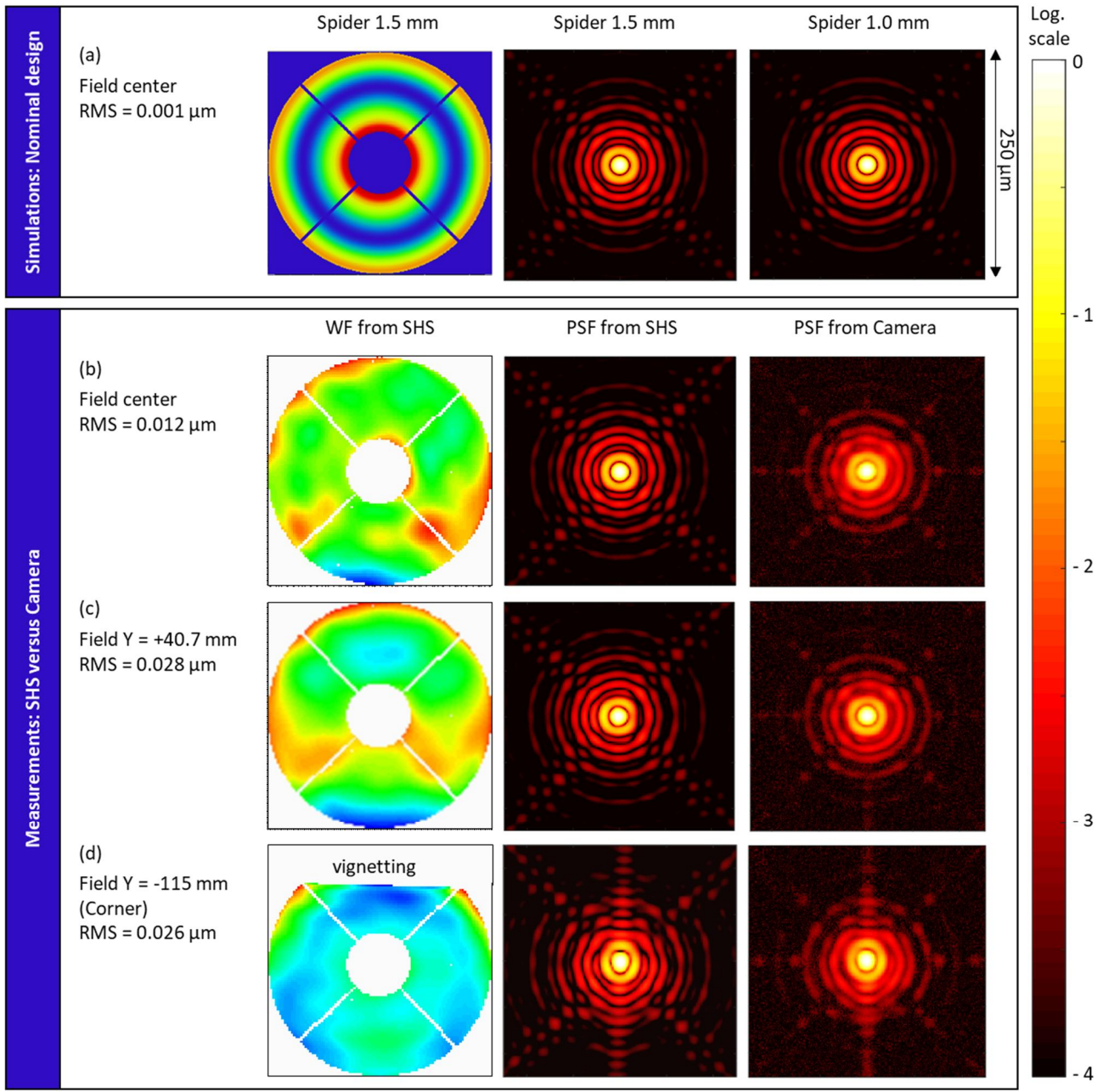

Figure 23. Row (a): Simulated WF and PSF for two different spider widths. Row (b-d): Comparison of PSF derived from the SHS measurement and PSF from the camera measurement for different FoV positions.

### 3.4 Determination of the Encircled Energy

The *encircled energy* (EE) is the proportion of the total energy of a light spot that lies within a circle, typically centred in the centroid. It is a measure of the spatial concentration of energy and commonly represented as a function of the radius R of the circle. The centroid is a simple and unique reference, which we use for our evaluations. Only for strongly asymmetric spots, other reference points may yield significantly steeper (and therefore 'better') EE(R) curves. The EE is important to determine the amount of energy from a point source (e.g. a star) that hits a single sensor pixel. It is a critical quantity that limits the visibility of faint objects.

#### 3.4.1 SHS based EE computation

The calculation of the SHS based EE is straightforward: We radially integrate the SHS based PSF with the help of a `Matlab` script. As the integration range is infinite, it has to be constrained for a numerical evaluation. We defined the maximum radius as $R = 7.6\,\lambda/NA = 146\,\mu m$. This value is adapted to the dynamic range of the 14-bit measurement camera, to keep the two approaches comparable. A larger radius would mainly increase the accumulated camera noise. (The SHS based measurements would also permit larger ranges.) The normalization is done with the ideal (diffraction-limited) EE at this point, which is $EE(146\,\mu m) = 0.9644$. The diffraction-limited case is calculated in `Matlab` based on the ideal pupil function (namely the radial integral of the square of the absolute value of the 2D Fourier transform of the ideal pupil function). See also Ref. [9] for details on EE evaluations. Again, we assume a constant pupil fill with central obscuration and a spider width of 1.5 mm, as discussed in section 3.3. The results are plotted in Figure 24 for the field centre and for the field position $Y = +40.7\,mm$, which, according to Figure 19, has one of the worst WFE RMS values. All other fields (apart from the corners, which cannot be reasonably compared due to the truncated WF) yield functions that are essentially in-between these two shapes. Note that the radius of the circle along the x-axis is represented in 'natural' units of $\lambda/NA$ and additionally at the top of the plot in units of the size of the NISP's H2RG$^{TM}$ detector pixels (18 $\mu m$ edge length) and in microns. These results are well reproducible with variations not much larger than the linewidths.

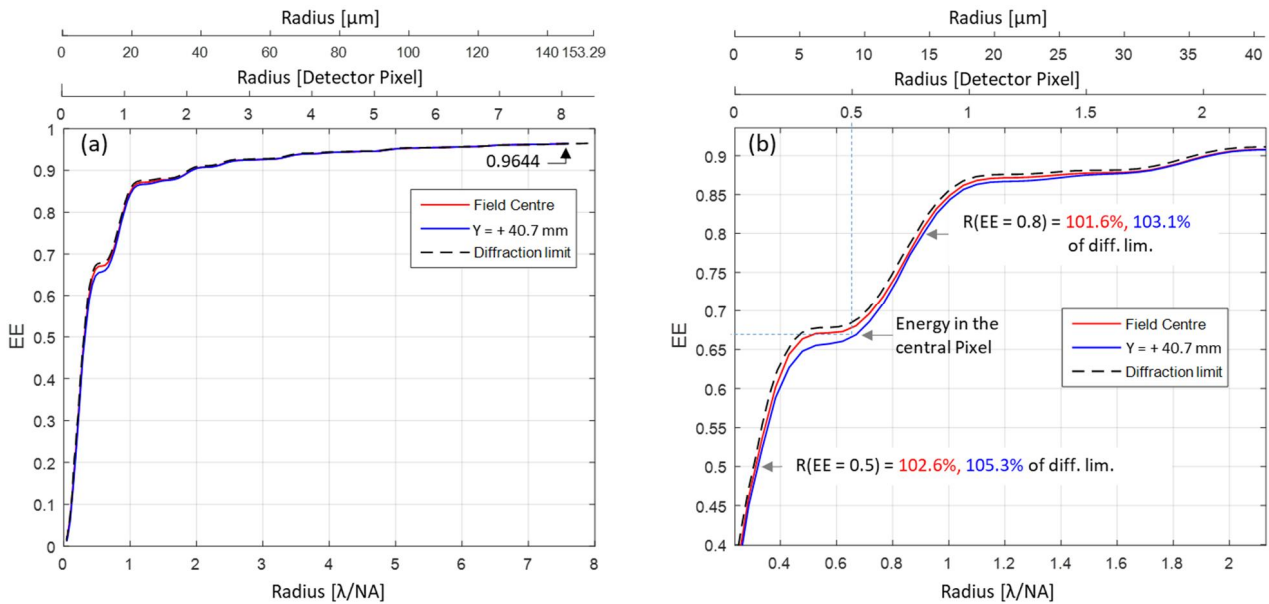

Figure 24. EE of the NI-OA @ $\lambda = 960$ nm based on SHS measurements for two different fields and comparison to the theoretical diffraction limit. (b) is an enlargement of (a). The radii for the 50% and 80% EE values are expressed as a percentage of the diffraction limited values. The x-axis is represented in 'natural' units of $\lambda/NA$ and in two additional units at the top of the plot.

#### 3.4.2 Camera-based EE computation

Compared to the SHS based approach, the camera-based EE computation has some drawbacks: The spot as detected on a camera sensor includes the impact of the source (spatial extent), the additional aberrations of the measurement optics and finally the actual effect of the 'unit under test'. A unique distinction between these different contributions is not possible, see also Ref. [9]. Another drawback comes with the fact that the representation of the PSF on the camera sensor requires a huge dynamic range of beyond 4 orders of magnitude. Large areas of the considered camera field have very low intensity and are thus affected by camera noise. Let us assess the impact of these drawbacks:

The difference between a mathematical point source with the corresponding constant pupil fill and our spatially extended fibre mode with MFD = 6.5 µm (= 0.28 $D_{Airy}$), which leads to a Gaussian pupil fill, is considered in Figure 25. While the impact to the PSF is rather unremarkable, there is some small, but recognisable effect in terms of the EE.

The measurement optics is well diffraction limited (section 2.8) and the (low) noise level of the cooled camera can additionally be reduced by averaging. These are comparatively good conditions for our camera-based EE measurements. Although the SHS based results are according to the mentioned drawbacks expected to be superior and more reliable, we want to further increase the confidence in our results by additionally evaluating the camera-based EE.

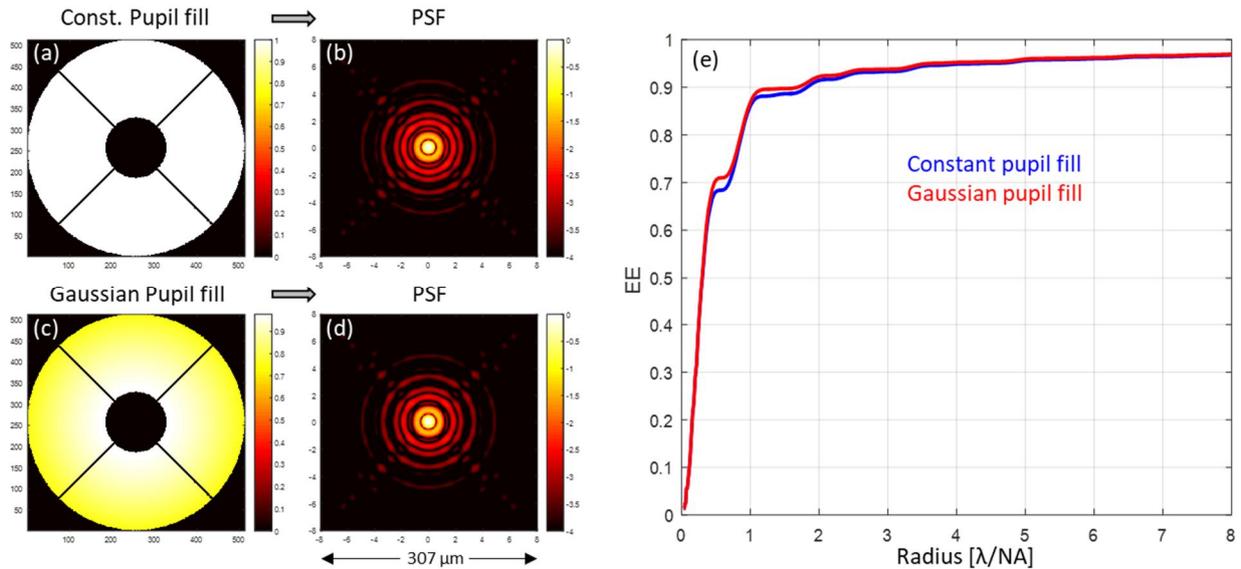

Figure 25. Theoretical calculation of the ideal PSF and the EE for a mathematical point source with the corresponding constant pupil fill (a) and for a spatially extended fibre mode with MFD = 6.5 µm (= 0.28 $D_{Airy}$), which leads to the Gaussian pupil fill in (c). While the impact to the PSF [(b) versus (d)] can hardly be observed, there is some visible effect in terms of the EE (e).

The camera-based EE is computed and normalized according to the SHS approach, however with one additional step: A tiny temporal drift of the dark camera count rate, which occurs probably in the read-out electronics, is corrected according to the procedure described in [9]. The diffraction-limited case assumes the physical spider width of 1.0 mm (in difference to the SHS approach with 1.5 mm).

Figure 26 shows the EE for three representative fields. All other fields (apart from the field-corners, which cannot be reasonably compared due to the truncated WF) lie essentially in-between these functions. Each measured function is an average over 40 EE computations from 40 single camera frames, which leads to a good reproducibility, although not as good as the SHS based results. The additional green curve is the result of the measurement *without* NI-OA according to the setup in Figure 8(b). For the camera measurements, it is *this* curve rather than the theoretical diffraction limited function that serves as a *reference* to indicate the impact of NI-OA. The aberrations due to the NI-OA are indeed marginal, in good accordance with the SHS based results.

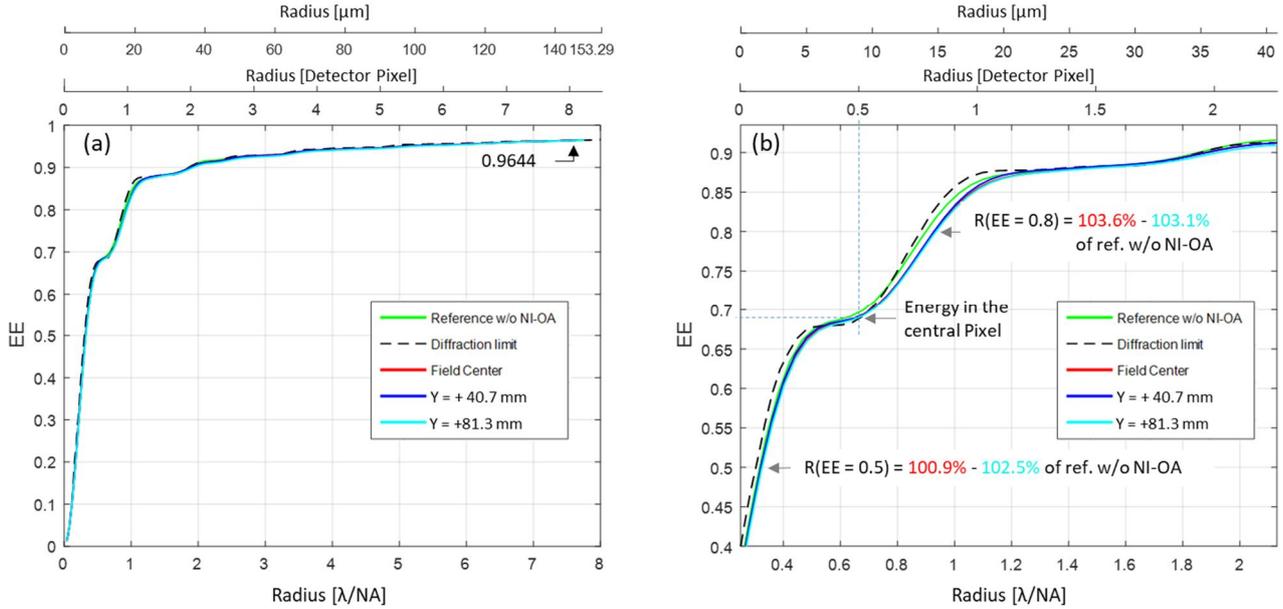

Figure 26. EE of the NI-OA @ λ = 960 nm. (b) is an enlargement of (a). In contrast to Figure 24, these results are based on camera PSF- instead of SHS measurements. The radii of 50% and 80% EE are given in percent of the measured values without NI-OA (green curve).

## 4. CONCLUSIONS

The near infrared optical assembly (NI-OA) of the *Euclid* space telescope is the largest civilian lens system ever launched into space. We developed a highly accurate setup to test its optical imaging properties under cryo-vacuum conditions. In the current paper, we discussed all critical components step by step: The object plane, the properties of the light source, the impact of the unavoidable cryostat window with regard to the optics as well as to the motion sequence of the hexapod, which guides the measurement devices to the field of view positions of our wide-angle device within single micron accuracy. Furthermore, we assessed the optical quality of the measurement devices itself, consisting of a Cassegrain-like two mirror telescope, additional lenses and the sensors.

The NI-OA's optical imaging quality was evaluated with two complementary approaches, namely a point-spread function- and a Shack-Hartmann sensor-based wavefront measurement. Both method have their characteristic pros and cons:

SHS-measurements provide in principle full information about all optical imaging properties. They are comparatively insensitive to the characteristics of the mapped object in terms of spatial extent, spectrum or coherence properties. Furthermore, a distinction between 'unit under test' and additional measurement optics is possible, if the setup allows a calibration measurement. On the other hand, the lateral resolution of the wavefront is restricted by the micro lens array of the SHS, which defines an upper cut-off frequency. In our case, this shows up as an incomplete resolution of the thin 'spider legs' and has an observable effect e.g. to the derived encircled energy.

In contrast, the camera-based direct PSF measurement resolves tiny structures in the pupil like the spiders completely. This approach is moreover quite sensitive to spatial extent, spectrum and coherence properties of the mapped object. That is sometimes desired and sometimes not. We use single mode optical fibres, because they can guide spatially highly concentrated light from the outside into the cryostat. The spatial extent of the field modes can however not be corrected by applying a deconvolution on the measured spot, because the modes are *spatially coherent*. Consequently, the mode field diameter must be small enough to be treated as a point source. Furthermore, a strict mathematical distinction between 'unit under test' and additional measurement optics is not possible and the huge required dynamic range of a PSF is challenging for a camera sensor and results in noticeable measurement noise in the areas with very low intensity. However, these properties make the PSF measurement sensitive to stray light, apodisation or 'ghost images' and provide not only confirmation but also supplementary information to the SHS results.

Our SHS-based measurements at λ = 960 nm distributed over the entire FoV show fantastic performance with typical wavefront error RMS values of λ/60 - λ/30. Even the worst case at one single field corner of λ/22 is far below the diffraction limit (defined as λ/14). In consideration of the discussed differences, both measurement approaches show good agreement at the level of the *PSF* and reproduce even slight substructures in logarithmic scale over 4 orders of magnitude as predicted. The *encircled energy*, derived from both approaches, confirms the very gratifying results. At λ = 960 nm, more than 2/3 of the energy of the PSF is focused on one pixel of the NISP's H2RG$^{TM}$ detector (18 μm edge length). The deviation of the *image curvature* from the nominal prediction, computed from the SHS measurements, is at least 4 times smaller than the Rayleigh length and therefore uncritical.

In summary, we simply do not expect any measurable reduction in the overall performance of the *Euclid's* near-infrared channel due to the NI-OA. Our results confirm the centring tolerances of the individual lenses of less than 5 μm measured during integration.

The entire NI-OA was manufactured and assembled twice. The second version is stored as the 'flight spare' (FS) model. The FS performance is only marginally worse than that of the flight model, showing the good control and repeatability of the manufacturing processes and the alignment methods.

## ACKNOWLEDGEMENTS

Many thanks to our partners in industry: Optocraft, Trioptics, Carl Zeiss Jena, Dioptics and OHB as well as to the Fraunhofer-Institut für Angewandte Optik und Feinmechanik (IOF).

We received numerous valuable comments and corrections from Jean-Gabriel Cuby and Peter Schneider who made intensive proofreading of the complete paper. Many thanks for that!

The Euclid Consortium acknowledges the European Space Agency and the support of a number of agencies and institutes that have supported the development of *Euclid*. A detailed complete list is available on the *Euclid* web site (http://www.euclid-ec.org). In particular the Academy of Finland, the Agenzia Spaziale Italiana, the Belgian Science Policy, the Canadian Euclid Consortium, the Centre National d'Etudes Spatiales, the Deutsches Zentrum für Luft- und Raumfahrt, the Danish Space Research Institute, the Fundação para a Ciência e a Tecnologia, the Ministerio de Economia y Competitividad, the National Aeronautics and Space Administration, the Netherlandse Onderzoekschool Voor Astronomie, the Norvegian Space Center, the Romanian Space Agency, the State Secretariat for Education, Research and Innovation (SERI) at the Swiss Space Office (SSO), and the United Kingdom Space Agency.

The MPE *Euclid* participation is supported by DLR under grant 50 QE 1101.